\documentclass{ar-1col}
\usepackage[numbers]{natbib}
\usepackage{hyperref} 
\usepackage{amssymb} 
\usepackage{amsmath} 
\usepackage{color}
\usepackage{todonotes}
\jname{Xxxx. Xxx. Xxx. Xxx.}
\jvol{AA}
\jyear{YYYY}
\doi{10.1146/((please add article doi))}

\begin{document}

% Page header
\markboth{Qin et al.}{The Hubbard model: A computational perspective}

% Title
\title{The Hubbard model: A computational perspective}

%Authors, affiliations address.
\author{Mingpu Qin$^1$, Thomas Sch{\"a}fer$^2$, Sabine Andergassen$^3$, Philippe Corboz$^4$, and Emanuel Gull$^5$
\affil{$^1$Key Laboratory of Artificial Structures and Quantum Control, School of Physics and Astronomy, Shanghai Jiao Tong University, Shanghai, China, 200240}
\affil{$^2$Max-Planck-Institut f{\"u}r Festk{\"o}rperforschung, Heisenbergstra{\ss}e 1, Stuttgart, Germany, 70569}
\affil{$^3$Institut f\"ur Theoretische Physik and Center for Quantum Science, Universit\"at T\"ubingen, Germany, 72076}
\affil{$^4$Institute for Theoretical Physics and Delta Institute for Theoretical Physics, University of Amsterdam, Amsterdam, The Netherlands, 1098 XH}
\affil{$^5$University of Michigan, Ann Arbor, MI, USA, 48109}}

%Abstract
\begin{abstract}
The Hubbard model is the simplest model of interacting fermions on a lattice and is of similar importance to correlated electron physics as the Ising model is to statistical mechanics or the fruit fly to biomedical science.
Despite its simplicity, the model exhibits an incredible wealth of phases, phase transitions, and exotic correlation phenomena.
While analytical methods have provided a qualitative description of the model in certain limits, 
numerical tools have shown impressive progress in achieving quantitative accurate results over the last years.
This article gives an introduction to the model, motivates common questions, and illustrates the progress that has been achieved over the last years in revealing various aspects of the correlation physics of the model.
\end{abstract}

%Keywords, etc.
\maketitle

%Table of Contents
\tableofcontents

% Heading 1
\section{Introduction}
\label{sec:introduction}
The Hubbard model is one of the simplest models of interacting fermions on a lattice. The model describes a fermion system with hopping term $t$ and interaction strength $U$ with Hamiltonian
\begin{align}
H = -t \sum_{\langle ij\rangle\sigma} \left(\hat c_{i\sigma}^\dagger \hat c_{j\sigma} + h.c.\right) +U \sum_i \hat n_{i\uparrow} \hat n_{i\downarrow}
\end{align}
where $i$ and $j$ denote sites on a lattice; $\sigma=\uparrow,\downarrow$ enumerates two spin species, $\hat c$ and $\hat c^\dagger$ annihilate and create particles, and $\hat n=\hat c^\dagger \hat c$. Sometimes next-nearest neighbor hoppings $t'$ are also included. Physical observables and phase diagrams are typically examined as a function of temperature, chemical potential (or, correspondingly, density), or (staggered) magnetic field.
In this form, the model goes back to papers by Hubbard \cite{Hubbard63}, Kanamori \cite{Kanamori63}, and Gutzwiller \cite{Gutzwiller63}. As already noted by these authors, the model is a sketch of nature in that it emphasizes electron correlation physics caused by local interactions in a single orbital, while phenomena due to non-local interactions, band structure, or effects between multiple orbitals are not directly contained.

Despite these radical simplifications, the model has proven itself as a powerful tool for investigating correlated electron physics. On one hand, its relevance to cuprate physics has provided an early motivation for studying the model's phase diagrams and ground states. On the other hand, its simplicity has made it an ideal target for early quantum simulators, where many-body phenomena can be investigated without the complication of many of the effects present in realistic condensed matter systems.
 
Theoretically, the presence of metallic, insulating, ferro- and antiferromagnetic, superconducting, and charge-ordered phases in a model with very few parameters has proven an appealing testbed for new analytical methods. However, it became apparent early on that the standard analytical toolkit of condensed matter theory was insufficient to describe this rich physics to the desired accuracy, and that sophisticated numerical methods would have to be used instead \cite{Gull15}. This led to the development of a wide range of numerical tools based on many different approximations and approaches \cite{LeBlanc15,Schaefer21A}, including diagonalization, diagrammatics, tensor network, variational, series expansion, 
Monte Carlo, and embedding methods. 
While different approaches often led to different answers in earlier years, the situation has significantly improved more recently. Thanks to algorithmic advances and an increase of computing power, several methods have started to reproduce consistent results, leading to  a growing consensus on various aspects of the Hubbard model.
 
\subsection{Purpose and structure of this article}
%%%%%%%%%%%%%%%%%%%%%%%%%%%
 \begin{figure}[t!]
   \centering
  \includegraphics[width=1.24\linewidth]{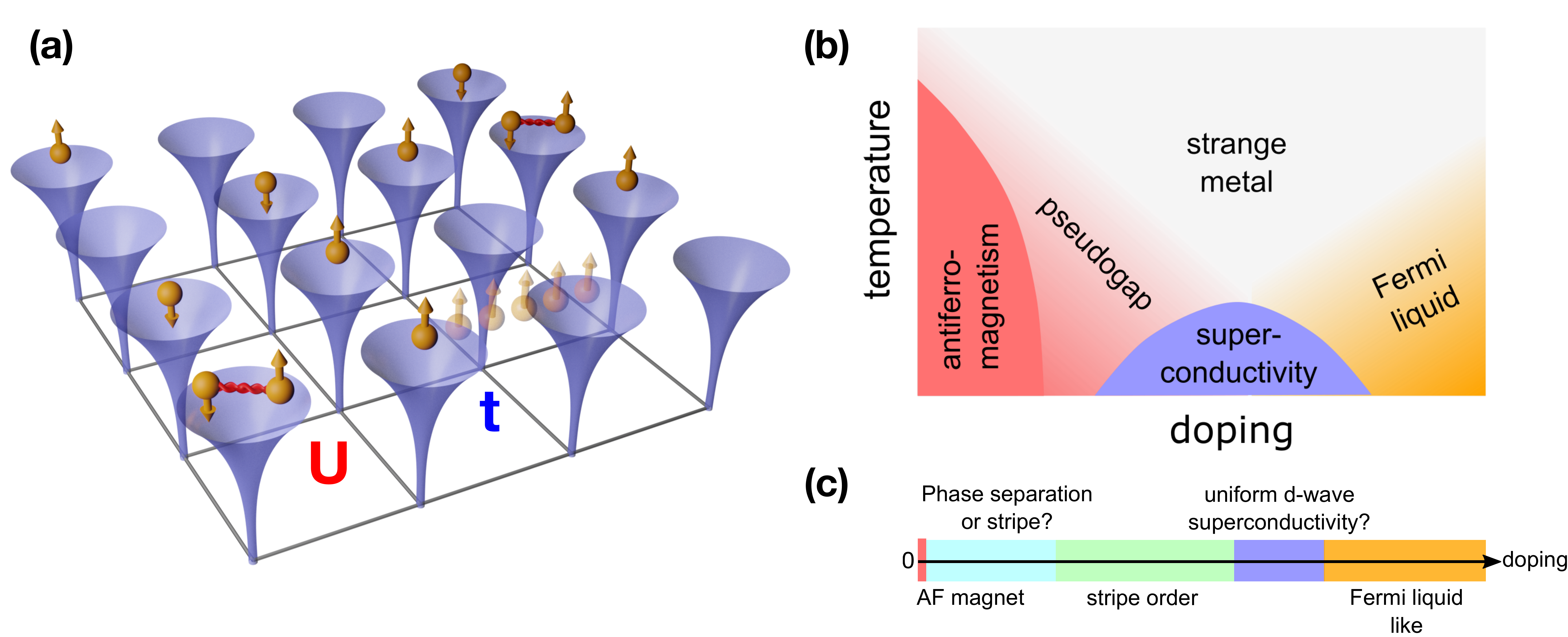}
   \caption{(a) Graphical representation of the Hubbard model. Two of many proposed phase diagrams of the model at intermediate interaction strength, at finite (b) and zero temperature (c). 
Note (see Sec.~\ref{sec:phase_diagram}) that (b) and (c) are mutually inconsistent, that in (b) charge ordered phases are missing and the precise location of phases and boundaries are hotly debated.
In (c) the ground state in the large doped region is Fermi liquid-like with an instability towards pairing through the Kohn-Luttinger effect \cite{Chubukov93}.    
   \label{fig:experiments}}
 \end{figure}
 %%%%%%%%%%%%%%%%%%%%%%%%%%%
In this article we review the recent progress in solving the Hubbard model from a computational perspective. We highlight results in which consensus has been reached among several numerical approaches and identify open challenges and their prospect of being resolved in the future.

Countless papers have been published on the Hubbard model over the last $30$ years. Inevitably, our list of references and our selection of topics can therefore not cover all of the important developments in this field. A good starting point for references to additional numerical works is given by earlier reviews \cite{Dagotto94,Bulut02,Maier05A,Tremblay06,Scalapino07,Scalapino12}. An analytical perspective is provided in Ref.~\cite{Arovas21} in the same issue. 

Our main focus is the single-band Hubbard model on the two-dimensional square lattice with interactions ranging from weak to intermediate and strong coupling. We also make connections to the $t$-$J$ model and the Hubbard model in three dimensions. In the remainder of this introduction we define what we mean by `solving' the model and discuss its connection to experiments. 
We will discuss the model in various parameter regimes, including broadly the `weak coupling' regime, which for the purpose of this article denotes interaction strengths $U/t \lesssim 4$ (rather than the mathematical limit where weak coupling perturbation theory is exact), the `intermediate-coupling' regime from $2 \lesssim U/t \lesssim 6$, and the strong coupling regime for $U/t \gtrsim 6 $ (which includes regions outside the applicability of an infinite-$U$ expansion).\begin{marginnote}
\entry{$t$-$J$ model}{an effective model of the Hubbard model in the large $U/t$ limit. At half-filling it reduces to the Heisenberg model.}
\end{marginnote}

We then review computational progress at half-filling (Sec.~\ref{sec:half-filling}), before we consider the doped case at weak (Sec.~\ref{sec:weak}) and intermediate to strong coupling (Sec.~\ref{sec:strong}). Finally, we review the progress towards the simulation of experimental probes (Sec.~\ref{sec:experimental}). We only briefly discuss other lattice geometries, attractive interactions, multi-orbital systems, non-local interactions, and extensions for non-equilibrium and driven systems (Sec.~\ref{sec:generalizations}). We conclude by giving perspectives and stating open questions (Sec.~\ref{sec:conclusions}).
To limit the scope of this article, we will also not discuss the bosonic version of this model, which has been studied extensively with cold atomic gas systems.

\subsection{`Solving' the Hubbard model}
Despite its simple Hamiltonian,  exact solutions of the Hubbard model only exist in special circumstances. 
Therefore, as a practical definition, we define as `solving' the Hubbard model the task of obtaining results for a physically interesting observable %in a parameter regime potentially  relevant for experiment, 
to an accuracy that is comparable to or better than what is obtained in comparison calculations or experiment. Corresponding to the richness of the physics of the model, `solving' the model can therefore have a very different meaning depending on the observable and parameter regime considered, and over time the questions of interest as well as the precision with which they have been answered have changed.

For instance, the question of whether there is $d_{x^2-y^2}$ superconductivity in the Hubbard model at $U/t\!\sim\!8$ and $1/8$ doping, which has been a major controversy for decades, has only been conclusively answered in recent years, when several methods confirmed that the ground state exhibits stripes but no superconductivity. While we consider this particular aspect as `solved', there remains room for improvement on the quantitative level, such as higher accuracy on the magnitude of the stripe order. 

 \begin{figure}[t!]
 \includegraphics[width=1.24\linewidth]{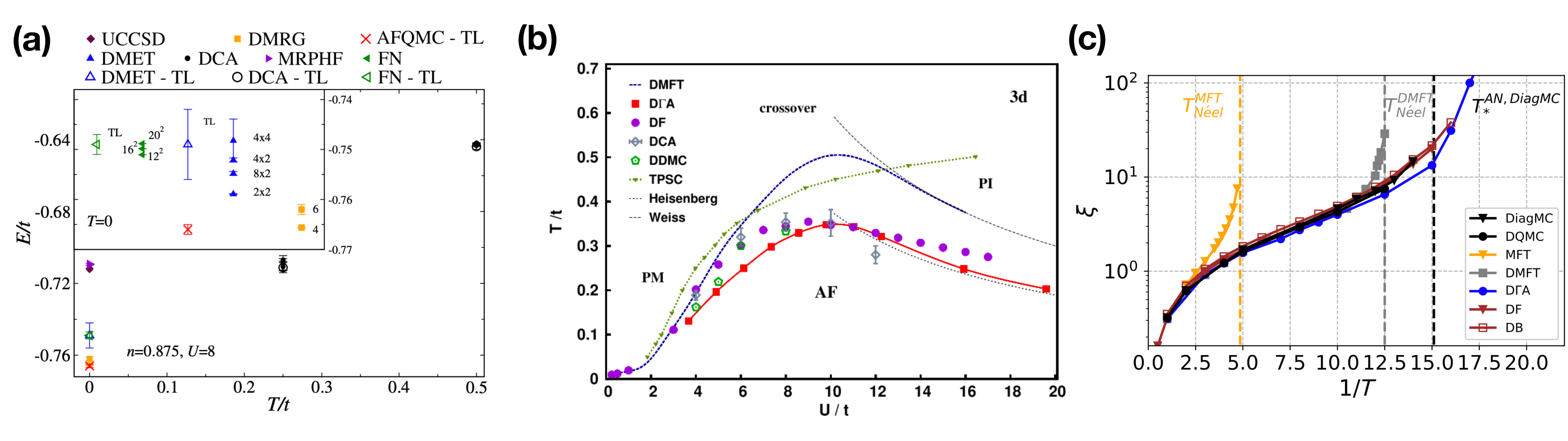}
   \caption{Consensus plots from benchmark studies of the Hubbard model.
   (a) Energy of the doped 2D square lattice model at $U/t\!=\!8$ as a function of temperature. Inset: Results for the ground - state energy (from \cite{LeBlanc15}). (b) N{\'e}el temperature of the half-filled model on a cubic lattice (central panel, reproduced with permission from APS \cite{Rohringer18} with work from \cite{Dare00,Staudt00,Rohringer11,Kent05,Hirschmeier15,Kozik13}). (c) Weak-coupling results for the temperature dependence of the magnetic correlation length $\xi$ in the half-filled model on a 2D square lattice ($U/t\!=\!2$, \cite{Schaefer21A}).
}
   \label{fig:bench}
 \end{figure}

It is frequently claimed, especially by works aiming to emulate the model in a quantum simulator or analog experiment, that the model is `exponentially hard' due to the `negative fermion sign problem' or the `exponential scaling of the Hilbert space', and therefore not tractable in numerical studies. These statements are simplifications of the real challenge, which is that in order to obtain properties of the model, all methods, simulators, and emulators obtain a sequence of approximate solutions followed by an extrapolation to the exact limit. These may be finite-size extrapolations, extrapolations in the number of variational states kept, in diagram order, excitation level, final time for time-evolution, coherence time, and/or optical trap parameters. Many of these parameters are strongly linked to the physics occurring in a given parameter regime. Correspondingly, the quality and accuracy of all methods for the `solution' of the model strongly depends on the parameter regime studied. In light of this it has been fruitful to establish `multi-method' consensus solutions \cite{LeBlanc15,Schaefer21A}, defined as solutions obtained with two or more methods with different underlying approximations, where results for the same quantity at the same set of parameters agree within method uncertainties. Figure \ref{fig:bench} shows results from such benchmark projects.

Collaborative benchmark projects have shown the value of obtaining results for the same parameters with multiple techniques. Even though several of the methods available are `controlled' or `numerically exact' in the sense that they converge to the exact result upon extrapolation in one or several parameters, this extrapolation is often difficult to perform in practice, or the methods are applied outside of the regime where the control parameters are small. Additionally, even though several techniques are `uncontrolled' in the sense that there are residual errors that cannot be extrapolated away, those errors may not be large when compared to the corresponding extrapolation or statistical uncertainties in controlled methods, or the respective technique may be much better suited to a particular problem. 
\subsection{Connection of the model to experiments}
\subsubsection{Cuprate physics}
Much of the interest in the Hubbard model in the intermediate to strong coupling regime stems from its relevance to the cuprate high-temperature superconductors discovered in the 1980s. Similar to experiments on these materials, Hubbard model simulations on the two-dimensional square lattice find regions of parameter space with $d_{x^2-y^2}$ superconductivity, strong antiferromagnetic correlations, stripes, pseudogaps, Fermi liquid, and bad metallic behavior. Phase diagram lines and observables mirror many features of the experiment as a function of doping and temperature.
However, it is difficult to make a precise correspondence between cuprate compounds and the single-orbital Hubbard model.

Electronic structure simulations \cite{Andersen95,Pavarini01,Zheng18,Hirayama18,Hirayama19} and photoemission experiment \cite{Damascelli03} find bands with oxygen $p$ and copper $d$ character near the Fermi surface, necessitating a description with at least three orbitals \cite{Hybertsen89}. 
As Zhang and Rice argued \cite{Zhang88}, a superposition of oxygen and  copper bands then yields an effective single orbital model.  
The parameters $t$ and $t'$ can then be inferred by fitting the tight binding dispersion of a two-dimensional square lattice with next-nearest neighbor hopping to the ab-initio band structure, leading to  $t'/t$ ranging from $-0.1$ to  $-0.4$ for most hole-doped cuprates \cite{Pavarini01,Markiewicz05,Hirayama18,Hirayama19}. 
Parameters can also be fitted to experiment. Fits to the photoemission signal \cite{Damascelli03} at the Fermi energy $E_F$ result in  $t'$ values of the same order and interactions of $U/t\!\sim\!8$ \cite{Kim98}. 
Similarly, the interaction parameter can be extracted by fitting the spin-wave dispersion of the model to results from neutron experiments \cite{Coldea01}.

These fits are neither systematic nor controlled. Nevertheless, they all yield interaction parameters comparable to the bandwidth, $U/t\sim8$, and a small negative $t'/t$. A more accurate description of cuprate materials likely would include additional degrees of freedom such as multiple orbitals per site, non-local interactions, and additional non-local hopping terms.

Recent experiments on superconducting nickelate compounds have rekindled the interest in the high-temperature superconductivity aspect of the model \cite{Kitatani20}. We also note that certain superconducting organic charge transfer salts may be described by a Hubbard model \cite{Powell06}.

\subsubsection{Quantum emulators and quantum computers}
Due to the simplicity of the Hamiltonian, the richness of the physics, and the failure of analytical methods to provide adequate solutions, the fermionic Hubbard model is often chosen as a first target for analog quantum emulators \cite{Georgescu14}. Hubbard quantum emulators aim to mimic the properties of the Hamiltonian by performing experimental measurements on a system that emulates the Hamiltonian. A wide range of emulators have been proposed, including (for the fermion model) ultracold atoms \cite{Mazurenko17} and quantum dots \cite{Byrnes08,Singha11,Hensgens17}. While these systems do not reach the predictive power of numerical simulations in equilibrium \cite{LeBlanc15,Schaefer21A}, interesting results have been obtained for time-dependent phenomena \cite{Schneider12,White19}.

An alternative route to `quantum emulation' is taken by early quantum computers \cite{Somma02}. Here, one may hope that the simple interaction and hopping structure of the model may reduce complexity enough that quantum computers can provide solutions. While current calculations lag far behind the capabilities of classical computers \cite{Arute20}, results for Hubbard models in interesting parameter regimes will likely become available long before quantum computing solutions of materials or chemistry problems are possible \cite{Wecker15}.

\section{Phase diagram and phase boundaries}
\label{sec:phase_diagram}
Not much is known about the precise phase diagram of the Hubbard model in two dimensions.
While the ground-state phase diagram of the model at weak coupling is fairly well understood \cite{Deng15,Simkovic16}, the phase diagram in the intermediate-to-strong interaction limit is hotly debated. Determining precise phase boundaries is hindered by the fact that the divergent correlation length at continuous phase transitions complicates finite size scaling; that several of competing phases have large and distinct unit cells; that energetic differences of competing phases are often minute; and that areas of phase space with very different physical properties are separated by broad crossovers, rather than sharp or first-order transitions.
In addition, approximate methods tend to break symmetries and overemphasize ordered phases. For instance, while the Mermin-Wagner theorem \cite{Mermin66} forbids magnetic order at non-zero temperature, most finite-size methods predict spurious magnetic phases that only disappear slowly in the limit of very large systems \cite{Maier05B}.

Correspondingly, while a broad consensus exists on the phases and their approximate locations, the precise shape of the transition lines is often not known. Visual representations such as the one in Fig.~\ref{fig:experiments} therefore either correspond to the results within a given approximation, or are inspired by comparisons to simple theories or experimental target systems.

The presence of so many orders and phenomena has led to the term `intertwined orders' \cite{Fradkin15}, which is used in this context to express the fact that very many different orders and phenomena either coexist in the same phase, or are otherwise very close in energy or parameter space, making it difficult to pinpoint the exact location and properties of each phase.

In the following, we will discuss some of these phases and their locations in parameter space. Numerous additional and exotic phases, including nematic phases, have been proposed for the Hubbard model. Ref.~\cite{Arovas21} provides an overview and discusses some of these phases in more detail.

\section{The Hubbard model at half filling}
\label{sec:half-filling}
%%%%%%%%%%%%%%%%%%%%%%%%%%%
 %%%%%%%%%%%%%%%%%%%%%%%%%%%
\subsection{The two-dimensional model at half filling}
%\begin{marginnote}
%\entry{`Hubbard' compounds}{
%Several `real' materials consist of single orbitals near the Fermi surface that are well isolated from the remainder of the orbitals and show strong local interactions. Their low-energy physics should therefore mimic that of a Hubbard model. They may be used to emulate the model or, in turn, present ideal targets for Hubbard model simulations. Material candidates include Moir\'{e} superlattices of WSe$_2$ and WS$_2$ \cite{Tang20} as well as LiCuF$_3$ and NaCuF$_3$ \cite{Griffin16}.
%}\end{marginnote}
The fermion sign problem is absent in Monte Carlo simulations of the half-filled Hubbard model on bipartite lattices \cite{Hirsch85}. This makes efficient QMC simulations on large lattices possible, either by directly simulating finite size systems or by embedding cluster impurity problems with quantum embedding methods \cite{Maier05A}.\begin{marginnote}
\entry{QMC}{
quantum Monte Carlo
}\end{marginnote}Both results can then be extrapolated to the exact thermodynamic limit. Precise results are also available from other methods, see {\it e.g.}
Refs.~\cite{LeBlanc15,Schaefer21A} for an overview of many physical quantities from a wide range of numerical methods.
\subsubsection{Static quantities: energies, entropies, and spatial correlation functions} The main static quantities of interest are ground-state and finite-temperature expectation values of energies, entropies, as well as spin and charge correlation functions. These quantities have been computed with a multitude of methods, and we summarize just a small selection of the results available.

Early finite lattice QMC results for antiferromagnetic order parameters and energies for different $U$ are presented in \cite{Hirsch89,White89}, a thorough study of magnetic properties on large systems was performed in \cite{Varney09}, and a list of ground state energies from the diagonalization of $4 \times 4$ systems at various interaction strengths and for different electron numbers can be found in \cite{Dagotto92}. More recently, ground state energies, double occupancies and
antiferromagnetic order parameters for different interactions extrapolated to the thermodynamic limit were presented in \cite{Qin16}, calculated from a finite-size scaling analysis with twist averaged boundary conditions. This work also tabulates finite-size data from $4 \times 4$ to $16 \times 16$ systems for different boundary conditions. 

Short-range correlation functions at half filling were calculated by QMC in \cite{Qin17} and finite temperature results can be found in \cite{Varney09}. In \cite{Kim20}, spin and charge correlations at finite temperature for $U/t\!\le\!4$ are calculated in a complementary DiagMC approach.
Entropy and specific heat for the same approach (also for the doped case) can be found in \cite{Lenihan21}. In the weak-coupling regime, the magnetic correlation length $\xi$ grows exponentially when lowering the temperature, see right panel of Fig.~\ref{fig:bench} \cite{Chakravarty88,White89,Moreo91,Schafer15,Rohringer16,Schaefer21A}. Eventually, these long-range fluctuations are responsible for the destruction of magnetic ordering (Mermin-Wagner theorem \cite{Mermin66}) and the development of a weak-coupling pseudogap (see Secs.~\ref{sec:gap} and \ref{sec:weak_pseudogap}) when approaching the (AF ordered) ground state from finite $T$.
\begin{marginnote}
\entry{DiagMC}{
diagrammatic Monte Carlo \cite{Prokofev98}}
\entry{AF}{
antiferromagnetic}\end{marginnote}

\subsubsection{Dynamical quantities: spectral functions}
The calculation of dynamical ({\it i.e.} frequency-dependent) results is considerably more challenging than the calculation of static observables. 

Monte Carlo methods formulated in imaginary time or Matsubara frequency provide dynamical quantities, such as single and two-particle spectral functions, via analytic continuation \cite{Jarrell96} (see \cite{Hirsch85} for gap extraction at half filling).
This continuation is ill conditioned and introduces additional uncertainty into the spectra even at half filling, where the sign problem is absent. In contrast, tensor network methods usually rely on a real-time evolution of the Hamiltonian, which avoids continuation but is limited by the growth of the entanglement with time, requiring increasingly large bond dimension. Alternatively, excitation spectra can also be obtained based on a tensor network excitation ansatz~\cite{Damme21}.

In the half-filled Hubbard model on the square lattice, QMC results for single particle gaps at $U/t\!=\!4$ were calculated in \cite{Furukawa92,Assaad96}. In \cite{Vitali16}, single particle gaps for different interactions were obtained in the thermodynamic limit by a careful finite size analysis, establishing the presence of a gap for all interactions with an exponential decay in the small $U$ limit and linear scaling for large $U$ \cite{Vitali16}.

Complementary results are provided by tensor network methods. In Ref.~\cite{Yang16}, the spectral function was obtained using time dependent DMRG as a solver in cluster perturbation theory for clusters up to $2\times80$, which was found to be in agreement with earlier QMC results~\cite{Bulut94,Preuss95,Preuss97,Grober00}, but with a higher resolution. Time-dependent MPS calculations of spectral functions on 4-leg and 6-leg cylinders have so far only been obtained for the $t$-$J$ model~\cite{Bohrdt20}. In \cite{Damme21} results for the excitation gaps for $U/t=12$ was obtained based on an MPS excitation ansatz on 4-leg and 8-leg infinite cylinders.

\begin{marginnote}
\entry{DMRG}{
density matrix renormalization group~\cite{White92}
}
\entry{MPS}{
matrix product state~\cite{Schollwoeck11}
}
\entry{DMFT}{
dynamical mean-field theory \cite{Georges96}
}\end{marginnote}

Resolving the full frequency and momentum dependence of the spectral function in the Brillouin zone is a difficult outstanding problem. Results from cluster DMFT are available at selected momentum points or averaged over small areas in momentum space, see {\it e.g.}
    \cite{Lin10}, but extracting the detailed momentum-dependence and Fermi surface shape yields results that strongly depend on the periodization schemes employed \cite{Klett20}.

\subsubsection{Nature of the gap}
\label{sec:gap}
The ground state of the two-dimensional Hubbard model on a square lattice with $t'\!=\!0$ is insulating, {\it i.e.} has a charge gap for any interaction strength $U$ \cite{Hirsch85,White89,Schafer15,Vitali16,Simkovic20}.
AF N\'{e}el order \cite{Hirsch89,Varney09} persists for all interaction strengths.

In the strong interaction limit, the gap is of the Mott type \cite{Mott49} and the low energy physics is described by an effective
spin-$1/2$ Heisenberg model with coupling constant $J = 4t^2/U$ \cite{Anderson59}.
The origin of the gap in the weak interaction region was controversial, and
Ref.~\cite{Anderson97} argued that the gap in the weak interaction limit remains Mott.
A complementary view is provided by the Slater mechanism \cite{Slater51}, where
the gap is formed by the establishment of long-range N\'{e}el order due to the perfect nesting of the Fermi surface. 

A diagnostic is provided by the evolution of the potential and kinetic energies as a function of interaction and temperature. In a Mott system, the transition to the insulator is accompanied by a lowering of the kinetic energy, whereas the Slater mechanism lowers the potential energy \cite{Gull08,Rohringer16}.
While early DCA calculations \cite{Moukouri01} supported Mott, later Refs.~\cite{Gull08,Schafer15,Rohringer16,Kim20} found a decrease of the potential energy in the weak coupling region, supporting a Slater mechanism.\begin{marginnote}
\entry{DCA}{
dynamical cluster approximation \cite{Maier05A}}\end{marginnote}
Despite the differing origins of the gap in the weak and strong coupling region, 
no phase transition between the two regimes is observed, indicating a crossover near $U/t\!\approx\!4$ \cite{Vitali16,Kim20}, which agrees with the properties of the underlying ground state \cite{Borejsza04}.

The evolution of the gap size and AF correlation length with temperature is difficult to study in numerical simulations, even where
the sign problem is absent, as the spin correlation length grows exponentially when temperature is decreased. Rich physics is found in
this process: upon cooling towards the insulating AF ground-state, a sequence of crossovers between a high-temperature incoherent regime, an intermediate metallic regime with coherent quasiparticles, and a low-temperature insulating regime with an AF pseudogap \cite{Schafer15,Simkovic20,Kim20} is observed in D$\Gamma$A, QMC and DiagMC.\begin{marginnote}
\entry{D$\Gamma$A}{
dynamical vertex approximation \cite{Rohringer18}}\end{marginnote}Due to the increased scattering rate at low $T$, the inverse quasiparticle lifetime (extracted from Matsubara data) shows a minimum as a function of temperature \cite{Rohringer16,Schaefer21B}. Results for further observables from different methods at weak coupling can be found in \cite{Schaefer21A}.
	
\subsection{The half-filled model in 3D}
For all interaction strengths, the half-filled model with $t'\!=\!0$ in 3D exhibits a phase transition from a paramagnetic state at high $T$ to an AF insulator at low $T$. The maximum transition temperature of $T_N/t\!\sim\!0.35$ occurs near $U/t\!\sim\!8$, remains large in the interval of $7\!<\!U/t\!<\!10$, and decays rapidly both for weak and for strong coupling \cite{Scalettar89B,Staudt00,Kent05,Rohringer11,Rohringer18} (see central panel of Fig.~\ref{fig:bench}). In addition to the AF phase transition, the system also exhibits a crossover identified either by a change in the compressibility or in the density of states at the Fermi energy near $U/t\!\sim\!6-8$ \cite{Staudt00,Fuchs11B} between metallic (at weak coupling) and insulating (at strong coupling) behavior in the high-temperature phase.

Lattice Monte Carlo methods \cite{Blankenbecler81} and cluster quantum impurity solvers \cite{Gull11} do not suffer from a sign problem and accurate results for the system are available for all interaction strengths \cite{Kent05,Paiva11,Kozik13}. The accurate treatment of the continuous phase transition requires considerable care due to the divergent length scale of fluctuations at the phase transition, and the transition has therefore been used as a rigorous test for diagrammatic extensions of the DMFT \cite{Rohringer11,Hirschmeier15,Rohringer18} and for finite size effects in cluster methods \cite{Kent05,Fuchs11,Kozik13}.
Critical exponents have been studied with diagrammatic extensions of DMFT \cite{Rohringer11,Hirschmeier15,Schaefer17} and, within those approximations, are compatible with Heisenberg universality for all interaction strengths.

The three-dimensional system is of particular interest in the context of cold Fermi gases  \cite{Jaksch05,Esslinger10}, where cooling towards an ordered state in three dimensions proved less challenging than probing two-dimensional superconducting and antiferromagnetic correlations, mainly due to the high critical temperature of the model. Entropy, rather than temperature, is conserved in closed traps, and the highest critical entropy of $S\sim0.65$ lies near $U\sim 8t$  \cite{Fuchs11}. Entropies of $0.77$ were reached in \cite{Jordens10}; AF correlations were observed in \cite{Greif13,Hart15}, and lattice cold atom experiments have since steadily increased the accessible parameter space \cite{Ibarra20}. 

\section{The doped 2D Hubbard model at weak coupling}
\label{sec:weak}
\subsection{Magnetic and superconducting properties}
\label{sec:weak_magn}
The ground state phase diagram reveals a rich behavior as a function of the different parameters.
In addition to AF N\'{e}el order, magnetic order with generally incommensurate wave vectors away from the N\'eel point $(\pi,\pi)$ can be found away from half-filling in mean-field studies \cite{Schulz90,Dombre90,Fresard91,Igoshev10} and, including fluctuations, by expansions in the hole-density \cite{Shraiman89,Chubukov92,Chubukov95,Kotov04}.
Incommensurate magnetic order is also indicated by diverging interactions and susceptibilities %at incommensurate momenta 
in fRG flows \cite{Halboth00B,Husemann09,Metzner12}, where approximate solutions indicate robust magnetic order up to fairly high doping provided that superconductivity is suppressed \cite{Yamase16}.
\begin{marginnote}
\entry{fRG}{
functional renormalization group \cite{Metzner12}}\end{marginnote}

While the magnetic instability in the Hubbard model is reproduced already by conventional mean-field theory, pairing is fluctuation-driven and hence more difficult to capture. Simple qualitative arguments suggesting d-wave pairing driven by magnetic fluctuations were corroborated by the fluctuation exchange approximation \cite{Scalapino95}. Convincing evidence for superconductivity at weak and moderate coupling strengths has been established by self-consistent or renormalized perturbation expansions \cite{Bickers89,Neumayr03,Kyung03,Raghu10}, DiagMC \cite{Deng15,Simkovic19} and from fRG calculations \cite{Zanchi00,Halboth00B,Honerkamp01A,Honerkamp01B,Metzner12,Eberlein14}. With its unbiased treatment of all fluctuation channels on equal footing, the fRG confirmed earlier studies based on the summation of certain perturbative contributions \cite{Scalapino12},
finding d-wave superconductivity with a sizable gap for a finite next-nearest neighbor hopping amplitude \cite{Eberlein14} coexisting with N\'eel or incommensurate antiferromagnetism in a broad doping range \cite{Reiss07,Wang14,Yamase16}.
When the chemical potential approaches the van Hove singularity, different instabilities compete: besides spin density wave and pairing instabilities, ferromagnetism has been observed at moderate $|t'/t|$ \cite{Irkhin01,Hlubina97,Honerkamp01B,Katanin03,Neumayr03,Raghu10}.
At finite temperature, computing the KT transition temperature from the superfluid phase stiffness, a superconducting dome centered around optimal doping has been found \cite{Vilardi20}.\begin{marginnote}
\entry{KT}{
Kosterlitz-Thouless}\end{marginnote}Recently, fRG flows starting from the DMFT solution instead of the bare action (DMF$^2$RG) \cite{Taranto14,Vilardi19}
confirmed robust pairing with d-wave symmetry at strong coupling, driven by magnetic correlations at the edge of the antiferromagnetic regime. %\sabine{shift this to strong coupling section?}

\subsection{Weak-coupling pseudogap}
\label{sec:weak_pseudogap}
The term `pseudogap' refers to a momentum-dependent suppression of the single-particle spectral function near the Fermi energy. In contrast to the pseudogap originating from strong coupling effects (see Sec.~\ref{sec:pseudogap}), at weak to intermediate coupling the pseudogap is induced by long-range antiferromagnetic correlations \cite{Vilk97,Borejsza04,Kyung04,Wu18,Schaefer21A}. The gap opens when the correlation length exceeds the thermal de Broglie wavelength $\xi \gg v_{F} / (\pi T)$, where $v_{F}$ is the Fermi velocity \cite{Vilk97}. 
The phenomenon has been studied within the two-particle self-consistent approach approximation \cite{Vilk97}, the D$\Gamma$A \cite{Schafer15}, the dual-fermion approach \cite{vanLoon18}, the parquet approximation \cite{Eckhardt20}, the DiagMC \cite{Simkovic20}, and the fRG \cite{Hille20B}.

The underlying mechanism can be understood already from the second-order contribution of the self-energy: using the Ornstein-Zernike form for the spin susceptibility,
%which accounts for the long-range antiferromagnetic spin fluctuations, it leads %the Schwinger-Dyson equation 
it predicts a spectral gap for momenta close to the hot spots where the Fermi surface crosses the magnetic Brillouin zone boundary \cite{Vilk97,Wu17}.
Upon lowering the temperature, the gap opens first near $(\pi,0)$ (the so-called antinode), then spreads across the Fermi surface until also the region near $(\pi/2,\pi/2)$ (the so-called node) becomes insulating. %This nodal/antinodal dichotomy has been found to originate in incommensurate spin fluctuations contributing significantly to the antinodal self-energy but not to the nodal one, whereas commensurate spin fluctuations with the exact nesting vector $(\pi,\pi)$ equally affect both \cite{Krien20}. 
While a momentum-selective gap opening due to long-range antiferromagnetic spin fluctuations has been observed also at electron doping \cite{Kyung04,Senechal04,Tremblay06}, most studies focus on models with a finite next-nearest-neighbor hopping $t'$ which show a non-Fermi liquid behavior of the self-energy at hole doping.
In the weak-coupling regime, there is a natural connection between the Fermi surface topology and the coherence of low-energy quasiparticles: for a hole-like Fermi surface, the coherence of low-energy quasiparticles is suppressed at the hot spots. When the Fermi surface turns electron-like, increased quasiparticle coherence is restored all along the Fermi surface. The pseudogap only exists when the Fermi surface is hole-like. Hence, the Lifshitz transition from hole- to electron-like topology controls both the location of the self-energy singularities and the topological transition of the Fermi surface.

\section{The doped 2D Hubbard model at intermediate-to-strong coupling}
\label{sec:strong}
\subsection{Competition of low-energy ground states: uniform vs stripe states}
\label{sec:competition}

%general intro to uniform vs stripes
A striking feature of the doped 2D Hubbard model at strong coupling as well as of the $t$-$J$ model is that they exhibit several competing ground states which lie very close in energy. This includes, on the one hand, a uniform d-wave superconducting state (see Refs.~\cite{Dagotto94,Bulut02,Maier05A,Tremblay06,Scalapino07} for a review of early results) 
sometimes coexisting with antiferromagnetic order at low doping, on the other hand various inhomogeneous states in which charge and/or spin densities are modulated, called stripes (see Refs.~\cite{Vojta09,Fradkin15,Kloss16,Agterberg20} for reviews). Experimentally, stripe order was also observed in some cuprate materials~\cite{Tranquada95} in the
under-doped region. In early Hartree-Fock studies, stripes were found to be insulating with a filling of exactly one hole per unit length~\cite{Zaanen89, Poilblanc89}, %~\cite{Zaanen89, Poilblanc89, Schulz89, Machida89}
 whereas later it was found that stripes may also exhibit a different hole density~\cite{White98, White98b, White03} with  coexisting d-wave superconductivity~\cite{White99,Himeda02,White09,Corboz11,Corboz14,Zheng16}. A more complex variant of the stripe, called a PDW state~\cite{Agterberg20}, includes a $\pi$ phase shift in the superconducting order between neighboring stripes, such that d-wave superconductivity vanishes on average.\begin{marginnote}
\entry{PDW}{
pair density wave}
\entry{LBCO}{
Lanthanum barium copper oxide, the first discovered high-temperature superconductor and member of the cuprate family.}
\end{marginnote}The PDW state was proposed in Ref.~\cite{Berg07} as a possible explanation for the suppressed superconductivity observed in LBCO around $\delta=1/8$ doping~\cite{Li07}, and was found to be energetically very close to the stripe without phase shift~\cite{Himeda02,Raczkowski07,Corboz14}.

%paragraph on the competition and consensus for t'=0
Due to this strong competition it has been an open question for many years whether the ground state is a uniform d-wave or a stripe state (with possible coexisting superconductivity). Initially, the former was supported particularly by many VMC studies~\cite{Yokoyama88,Gros88,Giamarchi91,Sorella02,Eichenberger07,Misawa14,Tocchio16} and (cluster) DMFT studies~\cite{Lichtenstein00,Halboth00B,Maier05B,Capone06}.\begin{marginnote}
\entry{VMC}{
variational Monte Carlo}
\end{marginnote}A common opinion was that the stripes found with DMRG could be an artifact of the cylinder geometry, and that they would not be stable in isotropic, periodic systems (which are difficult to study with DMRG). This viewpoint was supported by a VMC study~\cite{Becca01} where it was shown that a lattice anisotropy indeed leads to (spin) stripe correlations, and also by Ref.~\cite{Hu12} where no stable stripe state was obtained in the $t$-$J$ model, even when adding a (period 4) stripe bias and combining VMC with fixed-node MC and Lanczos steps. On the other hand, stripe states were found with iPEPS in the 2D thermodynamic limit~\cite{Corboz11} in the $t$-$J$ model, which are energetically lower, but very close to the uniform state~\cite{Corboz14}, and also in a inhomogeneous DMFT approach for $U/t=8$~\cite{Peters14}.
\begin{marginnote}
\entry{iPEPS}{
infinite projected entangled-pair states~\cite{Jordan08}}
\entry{MC}{
Monte Carlo}
\entry{(CP-)AFQMC}{
(constrained-path) auxiliary-field quantum Monte Carlo~\cite{Zhang97}}
\entry{DMET}{
density matrix embedding theory~\cite{Knizia12}}
\end{marginnote}

\begin{figure}[]
  \centering
 \includegraphics[width=1.24\linewidth]{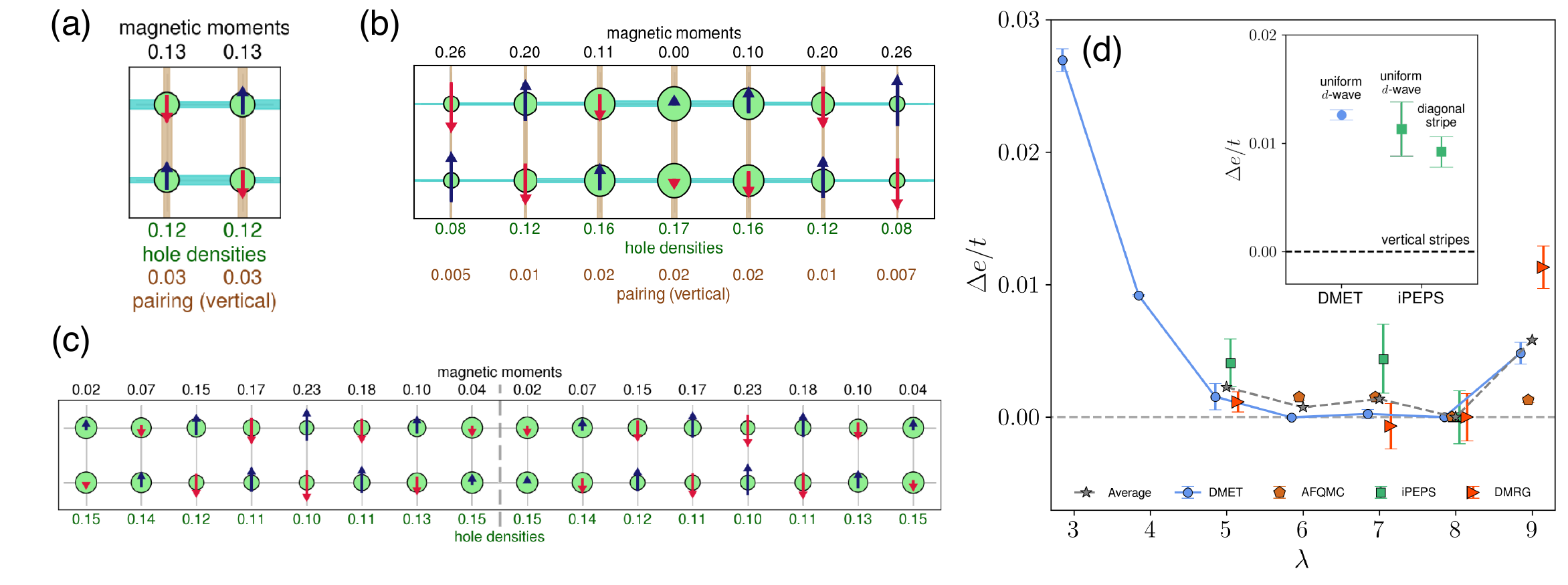}
  \caption{(a-c) Examples of competing states in the doped 2D Hubbard model for $U/t=8$ and $\delta=1/8$. The circle radius and the arrow length are proportional to the hole and spin densities, respectively, and the width of the colored bonds scales linearly with the local singlet pairing strength. (a) Uniform d-wave superconducting state with coexisting antiferromagnetic order (from iPEPS). (b) A stripe with charge and spin period 7 with coexisting d-wave superconductivity, and the typical $\pi$-phase shift in the antiferromagnetic order across the domain wall with maximal hole density (from iPEPS). (c) A period 8 stripe with exactly one hole per unit length per stripe, where superconductivity is entirely suppressed (from CP-AFQMC). Note that due to the $\pi$-phase shift in the AF order, the period of the spin order is doubled in stripes with an even charge period. (d)~Energies of stripes with different charge periods $\lambda$ and the uniform d-wave state (inset) relative to that of the $\lambda=8$ stripe obtained with different methods. All figures are taken from Ref.~\cite{Zheng17}.  
  }
  \label{fig:competingstates}
\end{figure}

More recently, based on a combined study with DMRG, CP-AFQMC, DMET, and iPEPS, consensus has been reached that the ground state of the doped 2D Hubbard model for $U/t=8$ ($t'=0$) and  $\delta=1/8$ doping is a stripe state with a charge period 8 without coexisting d-wave superconducting order~\cite{Zheng17}, shown in Fig.~\ref{fig:competingstates}(c). While stripes with periods 5-7 and coexisting d-wave superconductivity are energetically very close to the period 8 stripe, the uniform d-wave state is  higher in energy by $\approx 0.01t$, see Fig.~\ref{fig:competingstates}(d). The estimates of the energy for the period 8 stripe obtained from the four methods are in remarkably close agreement, lying in a range of $-0.767\pm0.004 t$. The period 8 stripe remains also stable in an extended interaction range ($U/t\sim6-12$), where the energy difference with respect to other stripes becomes larger with decreasing $U/t$, i.e. the strongest competition is found for $U/t=12$ (whether the stripe period shifts to lower periods at even larger $U/t$ is still an open question). 
Subsequently, the stripe ground state has been confirmed by VMC~\cite{Ido18,Tocchio19}, in determinant QMC~\cite{Huang18},  in other DMRG studies~\cite{Ehlers17,Huang18,Jiang19,Jiang20}, and by a variational AFQMC approach~\cite{Sorella21}.
The main reason why in previous VMC calculations~\cite{Hu12} stripe states were found to be higher in energy than uniform states is that only a period 4 stripe was considered, which is substantially higher in energy than the period 5-8 stripes~\cite{Zheng17,Tocchio19}. This resolves the previous discrepancy and also highlights the importance of using the correct unit cell in these calculations.

Moving away from $\delta=1/8$ doping, several approaches predicted that the stripe period decreases with increasing doping (see e.g. Refs.~\cite{Darmawan18,Ido18,Tocchio19,Sorella21}), with a tendency to stabilize insulating, fully-filled stripes of period $\lambda$ at doping $\delta=1/\lambda$ (for $U/t=8$, $t'=0$). For dopings in between these fully filled stripes, partially filled stripes with coexisting superconductivity have been found by VMC~\cite{Tocchio19}, whereas variational AFQMC predicted phase separation between insulating stripes~\cite{Sorella21}.  Based on a combined DMRG and CP-AFQMC study, it was concluded that superconductivity in the stripe ground state is absent  over an extended doping ($\delta\sim0.1-0.2$) and interaction ($U/t \sim 6-8$) range~\cite{Qin20}. At  low doping phase separation between the antiferromagnetic state at half filling and a partially filled stripe was found to occur around $\delta\approx 0.08$ for $U/t=8$~\cite{Tocchio19,Sorella21} ($\delta\approx 0.07$ for $U/t=10$~\cite{Darmawan18}), but a confirmation of this result with tensor network methods or CP-AFQMC is still lacking. 
(We note that the possibility of phase separation at low doping~\cite{Emery90} was already intensely studied many years ago, see  Refs.~\cite{Dagotto94,White00} for a review, but this concerned phase separation between the AF state and the uniform d-wave state).

The fate of the uniform d-wave state, i.e. under what conditions it is favored over stripe states, is an important question, but has not been investigated by all methods. The VMC study in Ref.~\cite{Tocchio19} predicted a stable uniform d-wave region in a doping range $0.20 \lesssim \delta  \lesssim 0.27$, for $U/t=8$ ($t'=0$), whereas a doping range $0.19 \lesssim \delta  \lesssim 0.22$ was found with variational AFQMC~\cite{Sorella21}.  Based on an approach combining VMC and a tree tensor network for $U/t=10$ ($t'=0$)~\cite{Darmawan18}, it was concluded that the uniform d-wave state gets stabilized for $0.17 \lesssim \delta  \lesssim 0.22$. 
A similar approach was used in Ref.~\cite{Ohgoe20} to study an ab-initio effective one-band model for Hg-based superconductors, where it was found that the long-range Coulomb interaction stabilizes the uniform d-wave state over an even larger doping region, $\delta \gtrsim 0.1$, with a competing period~4 stripe only remaining around  $\delta \sim 0.1$. These results suggest that, while the various competing states can already be observed on the level of the simplest Hubbard model, more accurate models with further-neighbor interactions and hoppings are required  to quantitatively reproduce the phase diagram of the cuprates. At the same time, to better resolve the tiny differences in energy for these competing states, more accurate methods need to be developed.

%%%%%%%%%%%%%%%%%%%%%%%%%%%
%%%%%%%%%%%%%%%%%%%%%%%%%%%

\subsubsection{Competition at intermediate $U/t$}
While the stripe ground state is now well established at strong coupling $U/t\ge6$, the relevance of stripes in the intermediate coupling regime is still not settled. At $U/t=4$, using DMRG and CP-AFQMC, the tendency for stripe formation was found to be weaker~\cite{Qin20} than at large coupling. A uniform d-wave ground state for $U/t\le4$ and large doping $\delta\ge0.3$ was predicted from DiagMC~\cite{Deng15,Simkovic19}, however, smaller dopings were not accessible due to convergence problems of the diagrammatic series. More work will be needed to accurately identify the crossover regime between uniform and stripes states.

%paragraph on finite t'
\subsubsection{Stripes at finite $t'$}
Stripe ground states have also been found upon including a negative next-nearest neighbor hopping $t'$ (corresponding to the hole-doped case) where the preferred stripe period shifts to smaller values with increasing strength of $|t'|/t$~\cite{Ido18,Huang18,Ponsioen19,Jiang19,Jiang20}. (A sufficiently large positive $t'/t \gtrsim 0.15$, in contrast, stabilizes the uniform d-wave state over the stripe state~\cite{Huang18,Ponsioen19}). 
Interestingly, a period 4 stripe, which is the typical  period found in experiments~\cite{Tranquada95}, is stabilized at $1/8$ doping over a wide range of $t'$, $0.16(4) < |t'|/t < 0.423(10)$~\cite{Ponsioen19} (a similar lower bound was found in Ref.~\cite{Ido18}), which includes the values that have been predicted as realistic parameters for different cuprate materials~\cite{Andersen95,Hirayama18,Hirayama19}. 
Both iPEPS~\cite{Ponsioen19} and VMC~\cite{Ido18} predict period 4 stripes without coexisting d-wave order at 1/8 doping. However, in contrast to VMC, iPEPS finds coexisting d-wave superconductivity in the period 5-7 stripes at 1/8 doping, and in the period 4 stripe at larger doping $0.14 \lesssim \delta  < 0.25$ (corresponding to a hole density per stripe unit length, $0.57 \lesssim \rho_l <1$; a similar range was also found for the period 5 stripe). 
Coexisting superconductivity in the period 4 stripe was found on a width-4 cylinder using DMRG \cite{Jiang19,Jiang20,Chung20} (also in the $t$-$J$ model~\cite{Dodaro17,Jiang18}), however, it was pointed out in Ref.~\cite{Chung20} that the pairing on the width-4 cylinder does not correspond to the ordinary d-wave order one would expect in the 2D limit, but to a "plaquette" d-wave pairing.  (Another difference between the width-4 cylinder DMRG and the iPEPS/VMC results is that in the former the spin order is not long ranged but only short ranged.) Clearly, the issue of coexistence or absence of superconductivity in stripe states and its nature (with or without phase shift between neighboring stripes) remains an important topic for future research.

\subsubsection{Competition in 3-band models of the cuprates}
\label{sec:competitionmultiband}
A competition between different low-energy states can also be found in 3-band Hubbard models with parameters relevant for the cuprates. Uniform d-wave superconducting states have been predicted in VMC~\cite{Yanagisawa01,Yanagisawa09,Weber14,Zegrodnik19, Biborski20,Zegrodnik20}, DMET~\cite{Cui20}, and cluster DMFT studies~\cite{Weber12}, while stripe states have been found in DMRG~\cite{White15} and determinant QMC (fluctuating stripes at finite temperature)~\cite{Huang17} studies. In the CP-AFQMC study in Ref.~\cite{Chiciak20}  stripes with spin and charge order were found for large values of the charge-transfer energy ($\Delta=4.4$ eV), whereas for small values ($\Delta=2.5$ eV) the charge order becomes weaker and a subtle competition between different spin orders was found. 
Stripes coexisting with superconductivity have been predicted in a VMC study~\cite{Yanagisawa09}, however, a confirmation of this result with more recent techniques is still lacking.

\subsection{Pseudogap}
\label{sec:pseudogap}

\begin{figure}[t]
   \includegraphics[width=1.24\textwidth]{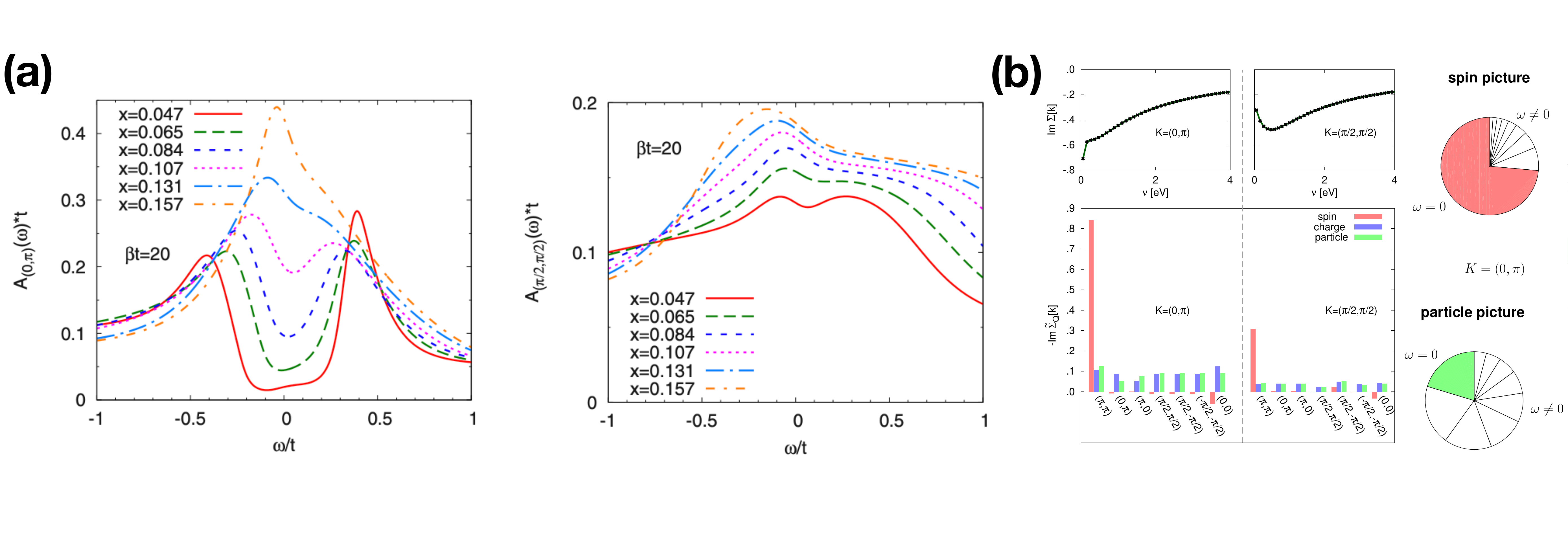}
   \caption{\label{fig:pg} (a) Many-body density of states from an eight-site dynamical cluster approximation calculation with $U/t\!=\!7$, $t'/t\!=\!-0.15$ and $T/t\!=\!0.05$ for various hole dopings $x$ at the antinode and node, respectively (reproduced with permission from APS \cite{Lin10}). (b) Fluctuation diagnostics of a DCA self-energy in the pseudogap regime, showing that short-range spin fluctuations are the origin of the pseudogap in the 2D Hubbard model (reproduced with permission from APS \cite{Gunnarsson15}).}
\end{figure}
In cuprate materials, the so-called pseudogap appears for dopings smaller than optimal doping and temperatures below $T^* \sim 300$K as a suppression of the density of states near the antinode $(\pi,0)$ in the Brillouin zone, but not near the node $(\pi/2,\pi/2)$. Originally it was identified as a suppression of the NMR Knight shift \cite{Alloul89}. However, signatures of the pseudogap are found in a wide variety of other probes \cite{Huefner08}, in particular angle-resolved photoemission spectroscopy \cite{Kanigel06,Shen05}.
Experimental correlation lengths are estimated to be on the order of a few lattice spacings \cite{Kastner98}.\begin{marginnote}
\entry{NMR}{
nuclear magnetic resonance}
\entry{CDMFT}{
cellular dynamical mean-field theory \cite{Maier05A}}
\end{marginnote}Therefore, the corresponding parameter regime in the Hubbard model is most often studied with cluster methods such as DCA and CDMFT, as they capture short-range correlation well and provide access to spectral functions.

\subsubsection{Results from cluster DMFT}
Pioneering results demonstrated that the inclusion of short-ranged correlations is sufficient to describe the suppression of the antinodal density of states in the single-particle spectral function upon lowering $T<T^*$ \cite{Maier00,Huscroft01,Macridin06}, see also left and middle panel of Fig.~\ref{fig:pg}. These DCA results were corroborated by $2\times 2$ CDMFT calculations with various impurity solvers, and their temperature, interaction, doping, and bandstructure ($t,t'$) dependence explored \cite{Parcollet04, Civelli05, Kyung06, Tremblay06, Zhang07, Liebsch09}. For instance, \cite{Civelli05} showed that, in addition to the strong renormalization of the Fermi surface at small dopings, there is a fundamental asymmetry of systems with negative next-nearest neighbor hopping $t'/t<0$ (resembling hole-doped cuprates) and such with $t'/t>0$ (electron doped): whereas the hole like Fermi surface curvature of the former is enhanced when approaching the insulating state for small dopings, the system is driven towards a nested Fermi surface in the latter. 

\subsubsection{Organizing principles: momentum-sector-selective metal-insulator transition, Widom line and Fermi-surface topology} Systematic DCA studies including variation of cluster sizes and doping \cite{Gull10} revealed a robust rationale for the perspective of the pseudogap regime in the $t-t'$ Hubbard model (where $t'$ is introduced to distinguish electron from hole doping) as a doping driven Mott transition (see middle panel of Fig.~\ref{fig:pg}): in the heavily doped regime, non-local correlations are small, yielding a nearly isotropic Fermi liquid (in the sense that the quasiparticle lifetimes and weights at the Fermi surface are nearly constant). 
Reducing the doping towards half-filling introduces a substantial differentiation in momentum space between quasiparticles at the node and at the antinode, such that the shorter lifetimes occur near the antinodal and the longest ones at the nodal point. The system however remains consistent with Fermi liquid theory at any point on the Fermi surface.
Reducing the doping further towards half filling establishes a gap near the antinode, while the node remains metallic. In cluster simulations, this is the momentum-dependent (`sector-selective') transition that characterizes the pseudogap in the Hubbard model \cite{Gull09}.
Reducing doping to half-filling leads to a phase transition to a Mott insulator with a substantially larger and isotropic gap. 
On the electron-doped side, the sector-selective regime is smaller or, if $t'$ is large enough, absent, and momentum-differentiation is weaker. It has been argued (see {\it e.g.} \cite{Parcollet04}) that the key ingredient for the pseudogap is the proximity to the Mott transition rather than long-range antiferromagnetic fluctuations, which are believed to be the gapping mechanism on the electron-doped side. A two-site minimal model exhibiting a momentum-selective transition is presented in \cite{Ferrero09A,Ferrero09}.

A different organizing principle has been suggested in $2\times 2$ CDMFT clusters \cite{Sordi12}. There, several thermodynamic and dynamic determination criteria for $T^*$ are found to collapse onto a line of thermodynamic anomalies that emanates from the critical endpoint of the Mott transition at finite doping. This is the so-called Widom line.

Conflicting interpretations exist on the influence of nesting and the van Hove singularity at strong coupling. While $8$-site DCA calculations find no substantial effect of the van Hove singularity on the pseudogap \cite{Gull09}, DCA and DQMC simulations in \cite{Wu18,Wu20} find substantial effects. The discrepancy remains unresolved.

\subsubsection{Insights into the origin of the pseudogap via fluctuation diagnostic approaches}
While observables such as the single-particle spectral function show signatures of the pseudogap, they do not easily reveal the underlying fluctuations that causes this feature. This analysis can be performed by examining the dominant fluctuations (charge, spin or pairing) and their change upon entering the pseudogap regime. This technique is known as  fluctuation diagnostics \cite{Gunnarsson15}. The approach relies on the Dyson-Schwinger equation of motion which relates the (one-particle) self-energy $\Sigma$ to the (two-particle) scattering vertex $F$. Expressing this relation in different fluctuation channels, their (momentum- or frequency resolved) contributions to the self-energy $\tilde{\Sigma}_{\mathbf{Q}/\Omega}$ can be determined. The right panel of Fig.~\ref{fig:pg} shows the self-energy and its fluctuation diagnostics calculated in DCA in a region of the 2D phase diagram where the Hubbard model exhibits a pseudogap. While the momentum contributions $\tilde{\Sigma}_{\mathbf{Q}}$ in the charge and particle-particle channel are uniformly distributed over the momentum vectors considered, in the spin channel there is a clear peak visible at $\mathbf{Q}=(\pi,\pi)$. In frequency, only in the spin channel the zero-frequency contribution $\tilde{\Sigma}_{\Omega=0}$ dominates, i.e. only the spin fluctuations are long-lived. From these observations one can conclude that well-defined short-ranged magnetic fluctuations lead to the opening of the pseudogap in the 2D Hubbard model. In \cite{Wu17} a fully momentum resolved fluctuation diagnostics has been performed with DiagMC data. The interplay of spin- and charge fluctuations for wide parameter ranges has been computed in DCA in \cite{Dong20}. A systematic review of different types and applications of fluctuation diagnostic techniques can be found in \cite{Schaefer21B}.

\subsection{Superconductivity}
The spontaneous emergence of superconducting pairing in a system with strong repulsive interactions has been one of the primary drivers for studying the model, and several earlier reviews \cite{Dagotto94,Bulut02,Maier05A,Tremblay06,Scalapino07} present the topic in detail. At weak coupling, a superconducting ground state (with several different superconducting symmetries) exists \cite{Metzner12,Raghu10,Deng15,Simkovic19}, see Sec.~\ref{sec:weak_magn}. However, as mentioned in section~\ref{sec:competition}, in a substantial parameter regime around $U/t\!\sim\!8$ and for moderate doping, the phase is preempted by various types of stripes. Nevertheless, superconductivity is very close in energy to those states, and minor changes in the model, such as additional non-local hoppings or interactions, may drive it to a superconducting state.
The investigation of this superconducting state and its relationship to the pseudogap, charge-ordered, and metallic states, therefore remains an important topic of active research.

The  superconducting phase boundary can be determined in two ways: upon cooling from the high-temperature disordered phase by tracking the divergence of the superconducting susceptibility, or upon heating from the low-T phase by investigating the disappearance of the superconducting order parameter.

With normal state calculations, early QMC results on finite-size clusters found $d_{x^2-y^2}$ as the most likely pairing symmetry in the Hubbard model for moderate doping \cite{White89B}. Large cluster simulations with extrapolations could be performed in DCA \cite{Maier05B} and the critical temperature determined, and \cite{Chen15} explored the parameter regimes most conducive to superconductivity.

Cluster DMFT simulations directly in the ordered superconducting state have been performed by several groups on four-site (see {\it e.g.} \cite{Lichtenstein00,Maier04,Haule07,Civelli09,Sordi12}) and eight-site dynamical mean-field clusters \cite{Gull13,Gull12}. These simulations allow to analyze the nature and properties of the superconducting state (including energetics, spectral functions, gap functions, and other response functions), in addition to determining the phase boundary. While considerable finite size effects remain, generic trends such as the energetics of the model or the evolution of response functions coincide with other methods and observations on cuprate materials.

Of considerable interest has been the pairing mechanism of the model. Several works \cite{Monthoux91,Scalapino95}, including studies with DCA \cite{Maier06,Maier06B,Maier07,Maier07B,Gull14}, found that AF fluctuations play the role of pairing ``glue" in the model. In \cite{Kitatani19} the influence of the dynamical vertex structure on the superconducting transition temperature has been analyzed within the D$\Gamma$A and particle-particle scattering processes have been identified as a one of the main oppressors of $T_c$.

\subsection{Bad metal}
The model at high temperature and intermediate doping is metallic, but transport properties are inconsistent with Fermi liquid behavior. The characterization of the temperature dependence of the resistivity, which is believed to be linear in temperature over a wide range of temperature, and of other transport properties in this regime has been a matter of recent attention. Results are available from various methods, including linked cluster expansion \cite{Perepelitsky16}, dynamical mean field theory \cite{Pruschke95,Deng13,Perepelitsky16,Cha20}, and lattice Monte Carlo \cite{Huang19}.

Calculating the resistivity is technically difficult. The interpretation of finite-temperature results on the Matsubara axis, such as those obtained by continuous-time QMC dynamical mean field solvers or lattice Monte Carlo methods, is complicated by the fact that analytical continuation is unreliable at high temperature, due to the distance of the Matsubara points from the real axis, and that the spacing of Matsubara point changes linearly in $T$, thereby introducing a systematic uncertainty into any temperature dependence.
Current real frequency formulations, such as those based on DMFT (using NRG \cite{Deng13} or ED \cite{Cha20} as impurity solver), neglect vertex corrections \cite{Vucicevic19}. Thus, while the regime at very high temperature is well understood, there is an intermediate temperature regime where currently no reliable calculations exist. Modern DiagMC methods formulated in real frequency \cite{Taheridehkordi20} promise a resolution of this problem.\begin{marginnote}
\entry{NRG}{
numerical renormalization group \cite{Bulla08}}
\entry{ED}{
exact diagonalization}
\end{marginnote}

\section{Towards the simulation of experimental probes}
\label{sec:experimental}
\subsection{Angle-resolved photoemission spectroscopy}
Much less is known about single- and two-particle excitations in the intermediate-to-strong coupling regime than about the energetics and the phase diagram of the model.

The spectral function is directly observable in photoemission experiment. Up to matrix elements, it corresponds to the imaginary part of the retarded real-frequency Green's function, which can be obtained from finite-temperature methods (working on the imaginary frequency/time axis) via analytic continuation \cite{Jarrell96,Fei21}. However, the continuation kernel is ill conditioned, and uncertainties or stochastic noise on the imaginary axis are amplified exponentially, leading to unreliable spectral functions especially at high temperature, away from zero frequency, and for bosonic quantities.
Thus, while features such as the existence of a gap or a quasiparticle peak are typically robust, there is considerable uncertainty in the numerical value of gap sizes or peak heights.

Cluster methods \cite{Maier05A} have been used to obtain spectral functions at select points in $k$-space. Interpolation methods for the Green's function, self-energy, or for cumulants \cite{Stanescu06} can then be employed to obtain continuous $k$-space Green's functions and extract quasiparticle weights or Fermi surfaces. Special care must be taken to disentangle interpolation artifacts from data. In three dimensions, results from cluster DMFT \cite{Fuchs11B} and lattice Monte Carlo \cite{Ulmke96} are available.

Several aspects of the spectral function can be obtained without resorting to analytical continuation. For instance, the imaginary time quantity $\beta G(\beta/2)$ converges to the density of states at the Fermi energy for $T\rightarrow0$ \cite{Trivedi95} and quasiparticle weights, gap sizes, or the location of Metal-to-insulator transitions are accessible from fits of the Matsubara self-energy \cite{Gull10,Georges96}.

\subsection{Optical conductivities}
Within linear response theory, the response of a model to a weak external field is described by a two-particle correlation function consisting of a product of single-particle Green's functions and its vertex correction \cite{Rohringer12}.

The response to an externally applied electric field corresponds to a current-current correlation function \cite{Basov05}. In the simulation of a layered two-dimensional system, the `c-axis' contribution is proportional to a product of two single-particle Green's functions and the inter-plane hopping, and no vertex corrections appear \cite{Ferrero10,Lin10}.  Interplane conductivities show a clear signature of the pseudogap opening with temperature and doping, consistent with experiments on the cuprates. In contrast, the evaluation of the in-plane conductivity requires vertex corrections \cite{Lin09,Lin10}, which only disappear in single-site \cite{Jarrell95} and four-site CDMFT \cite{Haule07B} due to symmetry. Results for conductivities are available from cluster DMFT at low $T$ and single-site DMFT and high-temperature series expansion \cite{Perepelitsky16} as well as Lanczos \cite{Vucicevic19} at high $T$. 
Cluster methods find a large insulating gap in the undoped system, a Drude peak with `mid-IR' feature upon doping, and a clear Drude peak in the metallic regime \cite{Lin10}.
\begin{marginnote}
\entry{IR}{
infrared}
\end{marginnote}

\subsection{Raman spectroscopy}
Electronic Raman spectroscopy \cite{LeTacon06,Devereaux07} is a versatile photon-in photon-out technique used to probe correlations in correlated electron systems. Depending on the polarization of the incoming and outgoing light, different areas of the Brillouin zone are probed. On a square two-dimensional lattice, two main light polarizations  of interest are B$_{1g}$ (sensitive to areas around $(\pi,\pi)$) and B$_{2g}$ (highlighting ($\pi/2,\pi/2$)). 

Simulated Raman results on the 2D model are available from cluster DMFT, with \cite{Lin12} and without \cite{Lin10,Sakai13} vertex corrections, as well as in the superconducting state \cite{Gull13B}. They show a  two-magnon Raman peak in the insulator, signatures of the pseudogap upon doping, and a temperature- and doping evolution generally consistent with experiment.
\subsection{Magnetic response}
Simulating the full momentum-and energy-dependent magnetic susceptibility or structure factor, such as it would be measured by neutron spectroscopy, is an important open problem. Lattice and quantum cluster methods do not have enough momentum resolution to capture the momentum resolution in an unbiased way, and results therefore come from approximate methods such as dual fermions \cite{LeBlanc19,Li20A}. These results can be directly compared to neutron spectroscopy on the cuprates \cite{Coldea01}.

Quantum cluster methods are better suited to simulating the local magnetic susceptibility, as it is measured {\it e.g.} in nuclear magnetic resonance. Results for the Knight shift, spin-echo decay time, and relaxation rates are presented in \cite{Chen17} and show the characteristic suppression of the NMR Knight shift in the pseudogap regime, along with a differentiation between `oxygen' and `copper' relaxation rates.

\section{Generalizations and extensions}
\label{sec:generalizations}
\subsection{The 2D Hubbard model on other lattices}

We so far discussed the Hubbard model on the square and cubic lattices.
The model displays additional interesting properties on lattices that are not square. In particular, the interplay between frustration
and correlation may lead to exotic magnetism at half filling, and doping holes may result in additional exotic electronic states.
\subsubsection*{Honeycomb lattice.}The model on the honeycomb lattice is studied mostly due to its connection to graphene. The lattice is bipartite
but has a  Fermi surface that is very different from that of the square lattice. At half filling, the Fermi surface shrinks to two Dirac points, which causes  AF order to appear only when $U$ is larger than a critical value $U_c$ \cite{Sorella92}. 
This makes the model an ideal playground to study the interaction-driven metal-insulator transition. At half filling, due to the absence of the sign problem,
the model can be efficiently studied with QMC. 
It is now well established that there is a direct phase transition from a Dirac semi-metal to an AF Mott insulator
at a critical $U_c / t \approx 3.8$ \cite{Sorella12,Assaad13} (and no intermediate spin liquid phase). The phase transition is in the Gross-Neveu-Yukawa universality class \cite{Assaad13,Otsuka16}. 
Away from half filling, much attention was paid to one quarter doping, where the Fermi-surface has a nesting feature.
Nesting in a Fermi surface usually triggers an instability towards
 an ordered state in the weak interaction region. Chiral $d+id$ superconducting order was found by several groups with a variety of methods \cite{Nandkishore12,Wang12,Gu13}.
\subsubsection*{Triangular lattice.} The triangular lattice is the simplest lattice with geometric frustration and may be relevant for certain organic materials \cite{Kanoda11}.
At half filling, $120$ degree magnetic order was confirmed by different methods \cite{Capriotti99,White07} in the Heisenberg model. The evolution of the phase diagram with $U$ at half-filling is still under debate (see Refs.~\cite{Szasz20,Tocchio20,Li20A} and references therein). 
In the most recent DMRG study on an infinite cylinder~\cite{Szasz20} a chiral spin liquid phase was found between the metallic phase at weak interaction and the $120$ degree magnetically ordered phase at strong interaction. However, the latest VMC study~\cite{Tocchio20} did not find evidence of an intermediate spin liquid phase, except upon adding a sufficiently large $t'$. More studies of large systems with high accuracy are needed to resolve this controversy.
 At finite temperatures a multi-method study of the metal-to-insulator crossover has been recently performed \cite{Wietek21}, where increased chiral correlations coexisting with stripy antiferromagnetic correlations at intermediate coupling strengths have also been found. Progress in computing frequency-dependent magnetic and charge susceptibilities has been achieved based on the ladder dual fermion approximation~\cite{Li20A}.
\subsubsection*{Kagome lattice.} The Kagome lattice is another commonly studied lattice with geometrical frustration. Unlike other lattices discussed here,
no magnetic order is found at half filling in the strongly interacting limit, and there is a growing consensus that the ground state is a quantum spin liquid (see \cite{Liao17} and references therein).  The Mott transition at half-filling was studied in \cite{Ohashi06} with cluster DMFT, where a critical coupling $U_c/t\sim 8.22$ was found. In \cite{Jiang17}, a Wigner crystal state was
found in the lightly doped $t$-$J$ model based on DMRG. The investigation of the physics at larger doping is still underway. In \cite{Kaufmann20}, a multi-method study with DQMC, DMFT and D$\Gamma$A investigated the magnetic correlations across the Mott transition.

\subsection{The attractive model}
The attractive ($U < 0$) Hubbard model is often used as a  model system for electronic superconductivity and superconducting phase transitions.  On bipartite lattices, the attractive interaction can be transformed to a repulsive one
with a partial particle-hole transformation: $\hat c_{i\uparrow}\rightarrow \hat d_{i\uparrow},\hat c_{i\downarrow}\rightarrow(-1)^{i}\hat d_{i\downarrow}^{\dagger}$.
With this transformation, the spin-balanced attractive Hubbard model at arbitrary filling is mapped to a
spin-imbalanced repulsive Hubbard model at half-filling.

QMC methods do not suffer from a sign problem here, and therefore accurate calculations of very large system sizes at low temperature are possible \cite{Scalettar89A}. 
At half filling, onsite s-wave pairing and charge density wave order at ($\pi, \pi$) coexist \cite{Hirsch85}, which correspond to the AF N\'{e}el order in
the spin-balanced repulsive Hubbard model at half filling. Away from half filling, only the onsite pairing survives and the model displays a KT phase transition at finite temperature. The KT transition temperature goes to zero as the system approaches
half filling \cite{Scalettar89A,Moreo91}, see \cite{Paiva04} for its accurate determination. Moreover, superconductivity can be significantly enhanced if the Fermi energy is close to a logarithmic Van Hove singularity in the density of states \cite{Hirsch86}.
The attractive Hubbard model remains sign problem free even with a Rashba spin-orbit coupling \cite{Shi16}.

\subsection{Multi-orbital models}
One may envisage an approach that gradually, in a `bottom-up' manner, adds more-and-more orbitals and interactions to the model until the full electronic structure Hamiltonian is recovered. 
A first step along this route leads to multi-orbital models with local Hubbard interactions. The number of possible models quickly grows and parameter choices become material specific. 
Cuprate physics has inspired the 3-orbital  Emery \cite{Emery87,Cui20} or p-d model, motivated by a downfolding of electronic structure to three, instead of one, orbitals, and the bilayer Hubbard model \cite{Maier11}. For a discussion of competing ground states in the three-orbital model see Sec.~\ref{sec:competitionmultiband}. 
Oxide perovskites have been studied in three-orbital generalizations of the Hubbard model. Typically, inter-orbital `Slater-Kanamori' terms that are not of the density-density type are also included \cite{Imada98}. These interactions are characterized by three parameters $U$, $U'$, and $J$. The model, most often studied with single-site multi-orbital DMFT, has given rise to the so-called `Hunds' metals \cite{Georges13}. The joint application of band downfolding techniques and treatment of the resulting low-energy multi-orbital model with DMFT (DFT+DMFT) is a frequently used strategy for the description of realistic materials \cite{Kotliar06}.
\begin{marginnote}
\entry{DFT}{
density functional theory}
\end{marginnote}

\subsection{Models with non-local interactions}
Instead of adding additional orbitals with local interactions, one may consider adding additional non-local interactions to the single-orbital Hubbard model. The simplest step along this route leads to the extended Hubbard model, where a repulsive nearest-neighbor density-density term $V$ is added to the local $U$. The extended Hubbard model is a paradigmatic model for charge order, as the system is driven to a charge ordered state as $V$ is increased and the temperature is lowered, leading to an interplay of charge order, correlation, antiferromagnetism, metallic, and Mott insulating behavior.

For this reason, the system has been studied within DCA \cite{Terletska17,Terletska18,Jiang18B,Paki19}, where these interactions are treated explicitly, as well as within the so-called dual boson approach \cite{Rohringer18,Stepanov18}, where they are treated in a diagrammatic expansion around DMFT.

\subsection{Non-equilibrium and driven system}
Real-time dependence of Hubbard models outside of linear response is a topic that has recently found much interest. Non-equilibrium can occur in many varieties, including quantum quenches (where parameters of the model are suddenly changed); the coupling to an external driving field (such as an electromagnetic field in a pump-probe setup); or the application of a voltage and the evaluation of currents in the short-time transient or long-time steady-state limit. In the long-time limit, there is a question of thermalization to a steady state or to equilibrium, and there are interesting questions on the destruction and reestablishment of ordered phases and of  non-equilibrium `phases'  \cite{Oka19}.

Reliable and generic numerical tools for these setups are mostly absent, and developing them is a field with great promise. Single-site and small cluster dynamical mean field calculations have been performed for most of these setups \cite{Aoki14}. Floquet systems (i.e. systems exposed to a periodic drive) have also been studied in cold atomic gas setups \cite{Messer18,Sandholzer19}.

\section{Conclusions and perspectives}
\label{sec:conclusions}
%3 paragraphs:
%[Summary paragraph with highlights]
In summary, thanks to substantial advances with different computational methods in recent years, controlled numerical solutions of the Hubbard model in various regions of the phase diagram have become available. This has led to a consensus on several aspects of the model. For example, the longstanding controversy over stripe states has been resolved, and it is now widely accepted that they are the ground state over an extended doping, interaction, and $t'$ range. Similarly, there is a consensus that the pseudogap at strong coupling is caused by strong short-range antiferromagnetic fluctuations.

%[Open problems]
Despite all of this progress, there  remain many open challenges, such as determining the precise locations of phase boundaries (in particular the parameter region with d-wave superconductivity), investigating the role of phase separation at low doping with more methods, and understanding the interplay of superconductivity and stripe order in partially filled stripes. 
Another open problem is the accurate study of the competition between stripe and uniform phases at finite temperature, which will be important to connect the current ground state results (mostly from wave function based methods) with the finite temperature results (predominantly from Green's function based approaches). 
More accurate techniques beyond single-site DMFT for transport calculations (e.g. fully vertex corrected resistivities) are needed to fully reveal the physics of the high-temperature metallic regime, which so far has remained largely unexplored by numerical approaches.
Similarly, more direct and controlled access to real-time evolution and real frequency observables is an essential future direction.
The physics of models with more orbitals and/or longer ranged interactions remains largely unexplored with accurate methods, and reliable algorithms are yet to be developed. Such methods will be essential to connect calculations on the single-band model to multi-band Hubbard models and electronic structure Hamiltonians without resorting to crude downfolding techniques.

%[Prospects to address these problems]
In view of these rapid advances, the prospects of addressing some of these issues in the near future seem promising. For example, tensor network simulations in 2D have recently been extended to finite temperature (see e.g.~\cite{Wietek20,Czarnik19,Chen21}), and already  been used to study the onset of stripe correlations at $\delta=1/16$ doping~\cite{Wietek20}. More finite temperature results in other parameter regimes as well as excitation spectra~\cite{Bohrdt20,Vanderstraeten19,Ponsioen20,Damme21} from tensor network approaches can be expected in future. Progress in computing dynamical quantities has also recently been achieved by dynamical variational Monte Carlo approaches~\cite{Ferrari19,Charlebois20}, with a first application to the 2D Hubbard model in Ref.~\cite{Charlebois20}. Moreover, diagrammatic approaches based on the two-particle vertex are being extended to real frequencies \cite{Kugler21}.

Finite-temperature methods have similarly made rapid progress. Recent years have seen the advent of large-cluster dynamical mean field studies that could be extrapolated to the thermodynamic limit in practice; extensions of the dynamical mean field theory that became more-and-more accurate, and the emergence of continuous-time and DiagMC methods that are orders of magnitude more powerful than previous techniques. 

The development of new, generally applicable numerical techniques which overcome the shortcomings of existing methods therefore remains the most urgent need for addressing the open issues. While this article did not focus on technical aspects of these calculations, most of the advances achieved over the last 15 years can directly be traced back to algorithmic and numerical progress.

%Disclosure
\section*{DISCLOSURE STATEMENT}
The authors are not aware of any affiliations, memberships, funding, or financial holdings that
might be perceived as affecting the objectivity of this review. 

% Acknowledgements
\section*{ACKNOWLEDGMENTS}
We acknowledge helpful input from F. Becca, P.~M. Bonetti, M. Ferrero, A. Georges, C. Hille, S. Kivelson, M. Klett, F. Kugler, J.~P.~F. LeBlanc, T. Maier, W. Metzner, B. Ponsioen, F. {\v S}imkovic, and A. Toschi. EG acknowledges support by NSF DMR 2001465 and the Simons Collaboration on the Many-Electron Problem. SA acknowledges financial support from the Deutsche Forschungsgemeinschaft (DFG) through Project No. AN 815/6-1. PC acknowledges funding from the European Research Council (ERC) under the European Union's Horizon 2020 research and innovation programme (grant agreement No 677061). 
\bibliographystyle{apsrmp}
\bibliography{review}

\end{document}